\documentclass[12pt,letterpaper]{article}

\def\be{\begin{equation}}
\def\ee{\end{equation}}
\def\bea{\begin{eqnarray}}
\def\eea{\end{eqnarray}}
\def\bean{\begin{eqnarray*}}
\def\eean{\end{eqnarray*}}
\def\bt{\begin{table}[ht]}
\def\et{\end{table}}
\def\ba{\begin{array}}
\def\ea{\end{array}}
\def\bc{\begin{center}}
\def\ec{\end{center}}
\def\mco{\multicolumn}
\def\rep#1{{\bf #1}}

\def\jrn#1#2#3#4{{#1} {\bf #2}, #3 (#4)}
\def\NPB{Nucl. Phys. B}
\def\PLB{Phys. Lett. B}
\def\PR{Phys. Rept.}
\def\PRD{Phys. Rev. D}
\def\PRL{Phys. Rev. Lett.}
\def\PTP{Prog. Theor. Phys.}

\def\JMP{J. Math. Phys.}

\begin{document}


\pagestyle{empty}

\rightline{UFIFT-HEP-03/6}

\vspace*{2cm}
\begin{center}
\LARGE{Family Hierarchy from Symmetry Breaking\\[20mm]}
\large{Fu-Sin~Ling\footnote{e-mail: fsling@phys.ufl.edu} and
Pierre Ramond\footnote{e-mail: ramond@phys.ufl.edu}\\[8mm]}
\it{Institute for Fundamental Theory\\
Department of Physics, University of Florida,\\ 
Gainesville, FL, 32611, USA\\[15mm]}
\large{\rm{Abstract}} 
\\[7mm]
\end{center}

\begin{center}
\begin{minipage}[h]{14cm}
We investigate symmetry breaking patterns from replicated
gauge groups which generate anomaly-free and
family-dependent $U(1)$ symmetries.
We discuss the extent to which these symmetries can explain the
observed hierarchies of fermion masses and mixings.


\end{minipage}
\end{center}
\newpage

\pagestyle{plain}
\section{Introduction}
The origin of fermion masses and mixings in the Standard Model 
is still as mysterious as when the first elementary particles 
were being discovered. Unlike the couplings of fermions to spin 
one particles which are now understood in terms of Yang-Mills 
interactions, Yukawa couplings still await such a level of understanding. 
The lack of a first principle explanation for their patterns have led 
theorists to devise elaborate schemes~\cite{IR,BDH,kingross,BS}, 
none of which (including our owns~\cite{FM,FS}) are particularly convincing.  
A less ambitious approach, couched in the language of low energy effective 
field theories, is that advocated by Froggatt and Nielsen~\cite{FN} (FN): 
the hierarchies of masses and mixings stem from higher dimensional 
operators which, when evaluated in the desired vacuum, yield effective 
Yukawa interactions of the right strengths. 
This approach organizes the dimensions of these operators in terms of 
hitherto unknown $U(1)$ charges. The result is that the suppression 
level of a particular Yukawa coupling is related to its Froggatt-Nielsen charge. 

Among ideas and schemes that involve additional gauge symmetries (flavor
or family symmetries) to explain this fermion hierarchy problem,
models with $U(1)$ symmetries have been shown to be self-consistent, 
anomaly-free, and experimentally testable~\cite{FM,FS,NW,ST,maekawa}.
In these schemes, the addition of a particular symmetry that accommodates
data is not derived from first principles; yet it could be very useful in 
hinting at its possible origin.

Of the  many versions of  models of this type in the physics literature, 
those with chiral Froggatt-Nielsen charges are particularly restrictive, since 
their anomalies must be cancelled. 
We have investigated FN models with several charges: one is family independent 
and anomalous; its anomaly is cancelled by a dimension-five Green-Schwarz term 
at the cut-off~\cite{GSM}, together with anomaly-free family 
dependent FN charges which are responsible for the interfamily hierarchies.

Specially daunting to these models has been the recent determination of the 
neutrino mass and mixing patterns.
While the existence of neutrinos masses and mixings is perfectly natural and 
expected, the recent determination of {\em two} large neutrino mixing angles 
poses further theoretical challenges~\cite{FS}.

In this paper, we limit our investigation to the generation of 
\emph{anomaly-free family-dependent} charges. 
We investigate the type of mechanisms capable of 
generating such symmetries.

Cancelling  a chiral anomaly is always done by adding new fermions. 
The easiest  is to add fermions of the opposite chirality, 
the route Nature chooses for QCD. 
The next to easiest is to add chiral fermions of different chiralities 
such that the result adds up to zero,  Nature's choice for the hypercharge anomaly. 
Today we understand the latter as embedding the Standard Model fermions inside 
representations of anomaly-free groups.

Central to anomaly cancellation are the groups. The anomaly free groups 
are well known: all Lie groups except $SU_n$ for $n>2$. 
On the other hand, anomalies can occur only if the representations are complex, 
and so we focus on anomaly-free groups with complex representations: 
the spinors of $SO(2n+6)$ and the complex representations of $E_6$. 
In fact the three families of Standard Model fermions fit remarkably well in 
a spinor of $SO(10)$ or a $\bf 27$ of $E_6$. 
However the extra charges carried by these groups are not family-dependent, 
although they are anomaly-free over these representations. 

To  find  anomaly-free, family-dependent charges, we assume, in the spirit of 
Ref.~\cite{FGNS}, that the gauge group at the Planck scale
is replicated. Taking one copy of the same group $G$ per family,
for three families, the fundamental gauge group will be of the form
\be
G_{Pl}~=~G \times G \times G
\ee
The group $G$ should be simple, anomaly-free and should contain the 
group of the Standard Model, $G_{SM} \equiv SU(3) \times SU(2) \times U(1)_Y$. 
The fundamental group $G_{Pl}$ therefore contains 
$G_{SM} \times G_{SM} \times G_{SM}$. 
It has to be reduced to the diagonal $G_{SM}$,
because Standard Model gauge interactions are flavor blind. 
Anomaly-free charges with opposite signs for different families can then be 
generated by order parameters which we call bi- or tri-chiral, 
to distinguish them from the usual bi-vector vacuum values. 

Symmetry breaking down to the Standard Model group proceeds in several steps. 
We assume that at some stage, there are $U(1)$ family-dependent symmetries 
which dictate the orders of magnitudes of the Yukawa couplings. 
In this paper, we restrict the discussion to the generation of anomaly-free 
family-dependent phase symmetries. 

The question of family hierarchy reduces to a search for plausible unifying 
structures and the way to break them. 
The qualitative features of fermion masses and mixings
are encoded in the underlying group structure and the breaking path.

We will start with some examples that show how these ideas can be 
realized before working our way towards a realistic scheme. 

\section{Simple examples}

In order to fix our notation, and introduce key concepts, we start this 
investigation with simple examples.

\subsubsection*{\underline{$SU(2) \times SU(2) \longrightarrow SU(2)$}}

We can first restrict the discussion to two families. 
The starting point is two copies of  the simplest non-Abelian group, namely
$$
SU(2)^\alpha ~\times~ SU(2)^\beta\ ,
$$
where $\alpha$ and $\beta$ label the two copies. 
The two fermion families fall into the representations
$$
\psi _1~ \sim~ (\rep{2},\rep{1})\ ,\qquad 
\psi _2~ \sim~ (\rep{1},\rep{2})\ .$$ 
The order parameter that describes the symmetry breaking is
taken to be a scalar field, transforming as a bi-fundamental representation 
$$
H~ \sim~ (\rep{2},\rep{2})
$$ 
A suitable vacuum expectation value (vev) of $H$ is able to 
trigger the breaking to the diagonal subgroup
$$
SU(2)^\alpha \times SU(2)^\beta ~\longrightarrow~ SU(2)^{\alpha + \beta}
$$
because the order parameter contains a singlet of the unbroken subgroup
$$
\rep{2} \otimes \rep{2} = \rep{1} \oplus \rep{3}
$$
After symmetry breaking, the fermions transform as doublets
under the remaining diagonal group. 
As long as the Higgs potential has no extra symmetry, 
there is no left-over phase symmetry. 
The two families are not distinguished by any family-dependent symmetry.  

\subsubsection*{\underline{$SU(3) \times SU(3) 
\longrightarrow SU(2) \times U(1)$}}

Our next example starts with the group
$$
SU(3)^\alpha ~\times~ SU(3)^\beta\ ,
$$
with  two fermion families transforming  as
$$
\psi _1~ \sim~ (\rep{3},\rep{1}) ~\equiv~ \rep{3}^\alpha \, ,
\qquad \psi _2~ \sim~  (\rep{1},\rep{3}) ~\equiv~ \rep{3}^\beta .
$$
We ignore for the moment the important question of anomalies. 
If the gauge group of the low-energy theory is the diagonal 
subgroup $SU(2)$, different inequivalent symmetry breakings are possible. 
Consider first an order parameter of the form 
$H_1 \sim (\rep{3},\overline{\rep{3}})$; we call it a {\em bi-vectorial} 
order parameter, since the two copies appear the same way up to a conjugation. 
A suitable {\em vev} will trigger the breaking
\be
SU(3)^\alpha \times SU(3)^\beta ~\longrightarrow~ 
SU(2)^{\alpha + \beta} \times U(1)_{V^\alpha + V^\beta}
\ee
where $V^{\alpha,\beta}$ is the Abelian factor in the embedding
$$
SU(3) ~\supset~ SU(2) \times U(1)_V
$$
The two fermion families are not  distinguished after symmetry breaking 
by the remaining $U(1)$ factor, since 
$$
\rep{3}^\alpha\ , \rep{3}^\beta ~\longrightarrow~ 
\rep{2}_1 \oplus \rep{1}_{-2}\ .
$$
In contrast,  consider the {\em bi-chiral} order parameter, 
$H_2 \sim (\rep{3},\rep{3})$, which can produce the breaking
\be
SU(3)^\alpha \times SU(3)^\beta ~\longrightarrow~ 
SU(2)^{\alpha + \beta} \times U(1)_{V^\alpha - V^\beta}
\ee
In this case, the extra Abelian symmetry acts as a family symmetry with 
opposite charges for the two families
\bean
\rep{3}^\alpha ~\longrightarrow~ \rep{2}_1 \oplus \rep{1}_{-2}\\
\rep{3}^\beta ~\longrightarrow~ \rep{2}_{-1} \oplus \rep{1}_2
\eean
We use this toy example to introduce a notation that enables us to catalog
in a systematic way all the possible singlet directions that
can be chosen by the scalar field during the symmetry breaking.
A singlet under $SU(2)^i$ will be notated as $\rep{1}^i_{v^i}$
with its charge $V^i$ in the subscript. 
A direction singlet under the diagonal subgroup of 
$SU(2)^i \times SU(2)^j$ but not under each group separately
will be notated by $\rep{1}^{i+j}_{(v^i+v^j,\,v^i-v^j)}$.
A straightforward calculation of the product
$\rep{3}^\alpha \otimes \overline{\rep{3}}^\beta$, applied to the bi-vectorial 
order parameter yields the following  $SU(2)^{\alpha+\beta}$ singlet 
\be
(\,\rep{3}^\alpha \,,\,\overline{\rep{3}}^\beta\,)~
\rightarrow~ (\,\rep{2}^\alpha _1 \oplus \rep{1}^\alpha _{-2}\,)
\otimes (\,\rep{2}^\beta _{-1} \oplus \rep{1}^\beta _2\,)
~\rightarrow ~\rep{1}^{\alpha+\beta}_{(0,2)}~ 
\oplus~ (\,\rep{1}^\alpha _{-2} \otimes \rep{1}^\beta _2\,)\ .
\ee 
The first singlet $\rep{1}^{\alpha+\beta}_{(0,2)}$ leaves the family-independent 
$U(1)_{V^\alpha + V^\beta}$ unbroken. 
The second singlet,  $\rep{1}^\alpha_{-2} \otimes \rep{1}^\beta_{-2}$, 
leaves the larger subgroup 
$SU(2)^\alpha\times SU(2)^\beta\times U(1)_{V^\alpha + V^\beta}$ invariant.

The same analysis applied to the bi-chiral order parameter yields
\be
(\,\rep{3}^\alpha \,,\,{\rep{3}}^\beta\,)~
\rightarrow~  (\,\rep{2}^\alpha _1 \oplus \rep{1}^\alpha _{-2}\,)
\otimes (\,\rep{2}^\beta _1 \oplus \rep{1}^\beta _{-2}\,)~
\rightarrow~ \rep{1}^{\alpha+\beta}_{(2,0)}~ 
\oplus~ (\,\rep{1}^\alpha _{-2} \otimes \rep{1}^\beta _{-2}\,)
\ee 
It is clear that $\rep{1}^{\alpha+\beta}_{(2,0)}$ now leaves the diagonal 
subgroup $SU(2)^{\alpha+\beta}$ and the family-dependent phase symmetry 
$U(1)_{V^\alpha - V^\beta}$ unbroken. 
Although the fermions are chiral, this simple model is not realistic 
since the starting group is anomalous. 

\subsubsection*{\underline{$G \times G 
\longrightarrow SU(3) \times U(1)$}}

It is more difficult to generate a diagonal $SU(3)$, starting with
two copies of a simple gauge group $G$. 
The maximal embeddings leading to a single $SU(3)$ are
\be
\ba{ccc}
SU(4) ~\supset~ SU(3) \times U(1) & & 
\rep{4}~=~\rep{3}_{-1} \oplus \rep{1}_{3}\\
Sp(6) ~\supset~ SU(3) \times U(1) & & 
\rep{6}~=~\rep{3}_1 \oplus \overline{\rep{3}}_{-1}\\
SO(8) ~\supset~ SU(3) & & \rep{8}~=~\rep{8}\\
SU(6) ~\supset~ SU(3) & & \rep{6}~=~\rep{6}\\
G_2 ~\supset~ SU(3) & & 
\rep{7}~=~\rep{3} \oplus \overline{\rep{3}} \oplus \rep{1}\\
E_6 ~\supset~ SU(3) & & \rep{27}~=~\rep{27}\\
E_7 ~\supset~ SU(3) & & 
\rep{56}~=~\rep{28} \oplus \overline{\rep{28}}
\ea
\ee 
Only the first two embeddings contain an extra $U(1)$ factor.
$Sp(6)$ is  vectorial under $SU(3)$, and this chain cannot lead to a model 
with chiral fermions. This leaves $SU(4)$, putting anomalies aside. 
To avoid  anomalies, one can embed $SU(4)$ into $SO(7)$
$$
SO(7) ~\supset~ SU(4)~; \qquad
\rep{7}~=~\rep{6} \oplus \rep{1}\ ,
$$
but then the  $\rep{7}$ of $SO(7)$ is vectorial, yielding vector-like fermions 
under $SU(3)$
$$
SO(7) ~\supset~ SU(4) ~\supset~ SU(3) \times U(1)~; \qquad
\rep{7}~=~\rep{3}_2 \oplus \overline{\rep{3}}_{-2} \oplus \rep{1}_0
$$
Because of $SU(3)$ triality, a bi-chiral order parameter does not contain an 
$SU(3)^{\alpha+\beta}$ singlet. This suggests we consider three copies

\subsubsection*{\underline{$G \times G\times G 
\longrightarrow SU(3) \times U(1)$}}

We start with three copies 
$$
SU(4)^\alpha~\times~SU(4)^\beta~\times~SU(4)^\gamma\ ,
$$ 
together with three fermion families 
$$
\psi _1~ \sim~ (\,\rep{4}\,,\,\rep{1}\,,\,\rep{1}\,)\ ,
\qquad  \psi _2~ \sim~ (\,\rep{1}\,,\,\rep{4}\,,\,\rep{1}\,)\ ,
\qquad \psi _3~ \sim~ (\,\rep{1}\,,\,\rep{1}\,,\,\rep{4}\,)\ .
$$
In this case,  the  {\em tri-chiral} order parameter 
$$
H ~\sim~ (\,\rep{4}\,,\,\rep{4}\,,\,\rep{4}\,)
$$
is capable of breaking to the diagonal $SU(3)$. Using the decomposition, 
$$
SU(4) ~\supset~ SU(3) \times U(1)_V~; \qquad 
\rep{4}~=~\rep{3}_1 \oplus \rep{1}{-3}\ ,
$$
and using $SU(3)$ triality,
$$
\rep{3} \otimes \rep{3} \otimes \rep{3} \supset \rep{1} \,
$$ 
we see that the breaking
$$
SU(4)^\alpha \times SU(4)^\beta \times SU(4)^\gamma ~\rightarrow~
SU(3)^{\alpha+\beta+\gamma} \times U(1)_{V_\alpha-V_\beta}
\times U(1)_{V_\beta-V_\gamma}
$$
is obtained if the order parameter takes a vev along the singlet 
$\rep{1}^{\alpha+\beta+\gamma}_{(3,0,0)}$, where the three subscripts 
refer to the sum of the charges and their two differences. 
Hence this order parameter produces two family-dependent phase symmetries. 
This yields the following fermions: three quarks, distinguished by their 
family charges, $(1,0)$, $(-1,1)$, and $(0,-1)$, together with three 
singlets of charges $(-3,0)$, $(3,-3)$, and $(0,3)$.  
This model is chiral, but  riddled with anomalies.  

\section{Towards a realistic scheme}

We learned from these simple but unrealistic  examples that a good candidate 
for a replicated  gauge group must have complex representations to describe 
chiral fermions, and must be anomaly-free.  
The candidate groups are then either orthogonal groups with complex spinor 
representations, or the exceptional group $E_6$~\cite{E6}. 
Here, we limit ourselves to studying  groups that appear in the sequence
\be
E_6 ~\supset~ SO(10) ~\supset~ SU(5) ~\supset~ 
SU(3) \times SU(2) \times U(1)_Y
\ee
The complex representation $\rep{27}$ of $E_6$ decomposes itself as
\be
\ba{ll}
E_6 ~\supset~ SO(10) \times U(1)_{V'} &
\rep{27} ~=~ \rep{16}_1 \oplus \rep{10}_{-2} \oplus \rep{1}_4\\
SO(10) ~\supset~ SU(5) \times U(1)_V &
\rep{16}~=~\rep{10}_1 \oplus \overline{\rep{5}}_{-3} \oplus \rep{1}_5\\
& \rep{10}~=~\rep{5}_{-2} \oplus \overline{\rep{5}}_{2}\\
SU(5) ~\supset~ SU(3) \times SU(2) \times U(1)_Y &
\rep{10}~=~(\rep{3},\rep{2})_{1/3} \oplus (\overline{\rep{3}},\rep{1})_{-4/3} 
\oplus (\rep{1},\rep{1})_2\\
&~~ \overline{\rep{5}}~=~(\rep{1},\rep{2})_{-1} \oplus (\overline{\rep{3}},\rep{1})_{2/3}
\ea
\label{eq:dec}
\ee

\subsubsection*{\underline{$SO(10) \times SO(10) \times SO(10)
\longrightarrow SU(5) \times U(1)$}}

We first consider  the case 
$$
SO(10)^\alpha \times SO(10)^\beta \times SO(10)^\gamma 
~\longrightarrow~ SU(5)^{\alpha+\beta+\gamma} \times U(1)\ .
$$
A tri-chiral order parameter 
$$
H_1 ~\sim~ (\rep{16},\rep{16},\rep{16})
$$ 
can trigger the desired breaking to the diagonal $SU(5)$, because the product 
$\rep{16} \otimes \rep{16} \otimes \rep{16}$ contains a $SU(5)$ singlet, namely
\be
\rep{16}^\alpha \otimes \rep{16}^\beta \otimes \rep{16}^\gamma ~\supset~
\rep{10}^\alpha _1 \otimes \overline{\rep{5}}^\beta _{-3} \otimes 
\overline{\rep{5}}^\gamma _{-3} ~\supset~ S_{1 \alpha} \equiv 
\rep{1}^{\alpha+\beta+\gamma}_{(-5,4,4)}
\label{S1}
\ee
where the $U(1)$ charges for the $SU(5)^{\alpha+\beta+\gamma}$ singlet 
$\rep{1}^{\alpha+\beta+\gamma}$ are given in subscript in the form 
$(v^\alpha+v^\beta+v^\gamma,v^\alpha-v^\beta,v^\alpha-v^\gamma)$. 
The product $\rep{16} \otimes \rep{16} \otimes \rep{16}$ also contains 
a singlet under $SU(5)^3$, as can be seen from the decomposition~(\ref{eq:dec}).
These singlets are listed in Table~\ref{so1}.

\bt
\caption{\label{so1} Singlets of $SU(5)$ in the product
$\rep{16}^\alpha \otimes \rep{16}^\beta \otimes \rep{16}^\gamma$ 
and associated broken directions $(a,b,c)$.
Singlets obtained by permutation of the indices $\alpha$,
$\beta$, $\gamma$ are omitted.}
\bc
\begin{tabular}{|c|c|c|}
\hline
&&\\
$SU(5)$ Singlet in & Unbroken & Broken\\
$\rep{16} \otimes \rep{16} \otimes \rep{16}$ & 
non-Abelian Group & Direction\\&&\\
\hline \hline &&\\
$\rep{1}^\alpha _5 \otimes \rep{1}^\beta _5 \otimes \rep{1}^\gamma _5$ & 
$SU(5)^\alpha \times SU(5)^\beta \times SU(5)^\gamma$ & $(1,1,1)$\\&&\\
\hline &&\\
$\rep{1}^{\alpha+\beta+\gamma}_{(-5,4,4)}$ &
$SU(5)^{\alpha + \beta + \gamma}$
& ${(1,-3,-3)}$\\&&\\
\hline
\end{tabular}
\ec
\et

Because a singlet direction carries three charges, it will always leave 
two independent $U(1)$ symmetries unbroken. If all the charges vanish, then
a third independent $U(1)$ is also unbroken.
If we associate a vector $\vec{v}=(a,b,c)$ in a three dimensional space 
to the linear combination $a V^\alpha+ b V^\beta + c V^\gamma$, then
for each singlet with at least one non-vanishing charge, 
the vectors $\vec{v}$ corresponding to the $U(1)$ symmetries left unbroken 
by this singlet span a (hyper-)plane.
We can therefore associate to such a singlet a vector perpendicular 
to this plane, and call it the broken direction. 

For example, the singlet $S_{1 \alpha}$ is invariant under any linear 
combination of $U(1)_{V^\beta-V^\gamma}$ and $U(1)_{3 V^\alpha+V^\beta}$.
Therefore, its broken direction is given by a vector perpendicular to
$(0,1,-1)$ and $(3,1,0)$, so we can take $\vec{v}_{1 \alpha}=(1,-3,-3)$.

Among all the possible linear combinations of $V^\alpha$, $V^\beta$ and
$V^\gamma$, the sum $V^\alpha + V^\beta + V^\gamma$ 
has the distinctive characteristic to be family-blind. 
It will be left unbroken by a singlet $S$ if the vector $(1,1,1)$ is
perpendicular the broken direction $\vec{v}_S$ associated to $S$.
In other words, the $U(1)$ symmetry corresponding to $\vec{v}_S$
has to be traceless over the family index. For example, 
the singlet $S_{1 \alpha}$ does not leave the combination
$V^\alpha + V^\beta + V^\gamma$ unbroken.

We also notice that the initial gauge group is invariant under a 
permutation of the indices $\alpha$, $\beta$, $\gamma$.
However, a singlet $S$ can spontaneously break this permutation symmetry,
corresponding to the group $P_3 \cong Z_6$, down to a smaller subgroup $D$.
All the singlets obtained from $S$ by a permutation of the indices
$\alpha$, $\beta$, $\gamma$ belong to the same conjugacy class
under the coset $Z_6/D$.

For example, the singlet $S_{1 \alpha}$ is invariant only under the
exchange $\beta \leftrightarrow \gamma$. In other words, it breaks
the symmetry $Z_6$ down to $D=Z_2$. Therefore, we can construct
two equivalent singlets $S_{1 \beta}$ and $S_{1 \gamma}$, obtained
from $S_{1 \alpha}$ by cyclic permutation of the indices
$\alpha$, $\beta$, $\gamma$
$$
S_{1 \beta} \equiv \rep{1}^{\alpha+\beta+\gamma}_{(-5,-4,0)}~~, \qquad
S_{1 \gamma} \equiv \rep{1}^{\alpha+\beta+\gamma}_{(-5,0,-4)}
$$
Their corresponding broken directions are $\vec{v}_{1 \beta}=(-3,1,-3)$
and $\vec{v}_{1 \gamma}=(-3,-3,1)$, and are again obtained by cyclic 
permutation of the indices $\alpha$, $\beta$, $\gamma$.

The appearance of a spontaneously broken discrete symmetry can 
be potentially harmful in the cosmoligical context, because
it leads to the formation of domains and domain walls.
However, if the scale at which the symmetry is broken is very high,
which is the case here, their presence will be washed away during
inflation.

We can now combine several singlets in order to narrow down the unbroken symmetry.
By this, we mean that the order parameter can yield a non zero vev
in more than one $SU(5)$ singlet. The $U(1)$ factors left over are found as 
the intersection of the unbroken spaces for each of the non-zero singlet.
For example, the vev $<H_1> = S_{1 \alpha} \oplus S_{1 \beta}$ 
will trigger the breaking
$$
SO(10)^\alpha \times SO(10)^\beta \times SO(10)^\gamma
~\longrightarrow~ SU(5)^{\alpha +\beta +\gamma} \times U(1)_{Y_F}
$$
with
\be
Y_F ~=~ V^\alpha +  V^\beta -\frac{2}{3} V^\gamma
\label{eq:firstsym}
\ee
This yields  three $SU(5)$ families 
$$
(\,\overline\rep{5}^{}_{-3}~, ~\rep{10}^{}_{~1}~, ~\rep{1}^{}_{~5}\,)
\ ;~~~~\qquad (\,\overline\rep{5}^{}_{-3}~, ~\rep{10}^{}_{~1}~, ~\rep{1}^{}_{~5}\,)
\ ;~~~~\qquad (\,\overline\rep{5}^{}_{~2}~, ~\rep{10}^{}_{-2/3}~, ~\rep{1}^{}_{-10/3}\,)\ .
$$
The subscript refers to  their $Y_F$ values. This family symmetry distinguishes 
the third family from the first two. 

Let us emphasize the question of anomalies. The Abelian charge $Y_F$ given by
Eq.~(\ref{eq:firstsym}) is our first example of an anomaly-free and family-dependent
symmetry. The cancellation of  anomalies is achieved by completing the $\rep{16}$,
and is ensured because $SO(10)$ is an anomaly-free group. 
Therefore, the inclusion of all three right-handed neutrinos is necessary. 
If we take the $SU(5)$ representations separately, the mixed anomaly
coefficients $(Y_F \, SU(5) \, SU(5))$ will not vanish, but the contributions
from $\rep{10}$, $\overline{\rep{5}}$ and $\rep{1}$ compensate each other.

The mixed anomalies will only vanish over each $SU(5)$ representation in the
case where $Y_F$ is traceless. This can be achieved by taking  
a tri-vectorial order parameter instead of a tri-chiral one
$$
H_2 ~\sim~ (\rep{45},\rep{45,\rep{45}})
$$
With the branching rule
\be
SO(10) ~\supset~ SU(5) \times U(1)_V~; \qquad
\rep{45} ~=~ \rep{24}_0 \oplus \rep{10}_{-4} \oplus 
\overline{\rep{10}}_4+\rep{1}_0
\label{eq:br}
\ee
we can obtain singlets of the type 
$\rep{1}^\alpha \otimes \rep{1}^\beta \otimes \rep{1}^\gamma$,
$\rep{1}^{\alpha+\beta} \otimes \rep{1}^\gamma$
and $\rep{1}^{\alpha+\beta+\gamma}$. They are listed in 
Table~\ref{so2}, together with their broken direction.

Notice that the adjoint representation has the peculiarity that its tensor
product with itself always contains a singlet and an adjoint 
representation. Therefore, in the case of $SO(10)$, we
have
\be
\rep{45}^\alpha \otimes \rep{45}^\beta ~\supset~ \rep{1}~; \qquad
\rep{45}^\alpha \otimes \rep{45}^\beta \otimes \rep{45}^\gamma
 ~\supset~ \rep{1}
\ee
These singlets have a zero charge under the diagonal $U(1)_V$,
$U(1)_{V^\alpha+V^\beta}$ and $U(1)_{V^\alpha+V^\beta+V^\gamma}$ 
respectively, but do not have a charge under the orthogonal 
combinations (and hence, these are broken).
They are of course also singlets under the respective
diagonal $SU(5)$. However, using the branching rule~(\ref{eq:br}),
we can actually construct singlets contained in the tensor 
product of adjoint representations of $SU(5)$, which
have well-defined (and zero) charges under the orthogonal linear
combinations of $V$'s. As mentioned previously, singlets for 
which all the charges vanish have no associated broken direction.
\bt
\caption{\label{so2} Singlets of $SU(5)$ in the product
$\rep{45}^\alpha \otimes \rep{45}^\beta \otimes \rep{45}^\gamma$ 
and associated broken directions.}
\bc
\begin{tabular}{|c|c|c|}
\hline
&&\\
$SU(5)$ Singlet in & Unbroken & Broken\\
$\rep{45} \otimes \rep{45} \otimes \rep{45}$ & 
non-Abelian Group & Direction\\&&\\
\hline \hline &&\\
$\rep{1}^\alpha _0 \otimes \rep{1}^\beta _0 \otimes \rep{1}^\gamma _0$ & 
$SU(5)^\alpha \times SU(5)^\beta \times SU(5)^\gamma$ & None\\&&\\
\hline &&\\
$\rep{1}^{\alpha+\beta}_{(0,0)} \otimes \rep{1}^\gamma _0$ &
$SU(5)^{\alpha + \beta} \times SU(5)^\gamma$
& None\\&&\\
$\rep{1}^{\alpha+\beta}_{(0,8)} \otimes \rep{1}^\gamma _0$ &
$SU(5)^{\alpha + \beta} \times SU(5)^\gamma$
& ${(1,-1,0)}$\\&&\\
\hline &&\\
$\rep{1}^{\alpha+\beta+\gamma}_{(0,0,0)}$ &
$SU(5)^{\alpha + \beta + \gamma}$
& None\\&&\\
$\rep{1}^{\alpha+\beta+\gamma}_{(0,8,0)}$ &
$SU(5)^{\alpha + \beta + \gamma}$
& $(1,-1,0)$\\&&\\
\hline
\end{tabular}
\ec
\et

The singlet $S_2=\rep{1}^{\alpha+\beta+\gamma}_{(0,8,0)}$ has the desired
features in order to give rise to a traceless family symmetry. 
The $U(1)$ corresponding to its broken direction is traceless,
therefore, it triggers the breaking
$$
SO(10)^\alpha \times SO(10)^\beta \times SO(10)^\gamma
~\longrightarrow~ SU(5)^{\alpha+\beta+\gamma} \times 
U(1)_X \times U(1)_{Y_F}
$$
with
$$
X~=~V^\alpha + V^\beta + V^\gamma \, , \qquad
Y_F~=~V^\alpha + V^\beta -2 V^\gamma
$$
The charge $X$ is family-blind while the charge $Y_F$ is traceless.
This yields the following three chiral families along with their $Y_F$ charges
$$
(\,\overline\rep{5}^{}_{-3}~, ~\rep{10}^{}_{~1}~, ~\rep{1}^{}_{~5}\,)
\ ;~~~~\qquad (\,\overline\rep{5}^{}_{-3}~, ~\rep{10}^{}_{~1}~, ~\rep{1}^{}_{~5}\,)
\ ;~~~~\qquad (\,\overline\rep{5}^{}_{~6}~, ~\rep{10}^{}_{-2}~, ~\rep{1}^{}_{-10}\,)\ .
$$

As emphasized earlier, we notice that the mixed anomaly coefficients 
$(Y_F \, SU(5) \, SU(5))$ between $SU(5)$ and the traceless family charge
$Y_F$ vanish over each $SU(5)$ representation. The charge $Y_F$ is said to be
non-anomalous in the language of Froggatt and Nielsen. 
In contrast, the anomaly coefficients $(X \, SU(5) \, SU(5))$ for a given
$SU(5)$ are necessarily different from zero, but are family-independent.
The charge $X$ is said to be anomalous. 

The anomaly coefficient $(Y_F \, X \, X)$ also vanishes because $Y_F$ is traceless.
Finally, the anomalies $(Y_F \, Y_F \, Y_F)$, $(X \, Y_F \, Y_F)$ and
$(X \, X \, X)$ can differ from zero for a given $SU(5)$ representation,
but they don't involve the Standard Model group or charges.
As before, a complete cancellation of the anomalies is achieved by
completing the $\rep{16}$ of $SO(10)$.

In the case that we have considered so far, we were able to construct
a family symmetry that is anomaly-free, and get three families of
chiral fermions with different family charges.
However, the breaking of $SO(10)$ down to $SU(5)$ lowers the rank
by only one unit, which does not leave enough room to build a family
symmetry that can accommodate the observed phenomenology of fermion masses.
This can be achieved by upgrading to $E_6$, as we are ready to see now.

\subsubsection*{\underline{$E_6 \times E_6 \times E_6
\longrightarrow SU(5) \times U(1)$}}

A further complication arises here, because $SU(5) \times U(1)$ is
not a maximal subalgebra of $E_6$. Two chains of maximal subalgebras
can lead from $E_6$ to $SU(5)$, namely (leaving aside the possible $U(1)$
factors)
\be
E_6 ~\supset~ SO(10) ~\supset~ SU(5)
\label{chain1}
\ee
which is the chain we already considered, or
\be
E_6 ~\supset~ SU(6) \times SU(2) ~\supset~ SU(5) \times SU(2)
\label{chain2}
\ee
In what follows, we will consider order parameters which are
tri-chiral, tri-vectorial, bi-chiral and bi-vectorial
\bea
\label{eq:hach}
H_1 ~\sim~ (\rep{27},\rep{27},\rep{27}) \nonumber \\
H_2 ~\sim~ (\rep{78},\rep{78},\rep{78}) \nonumber \\
H_3 ~\sim~ (\rep{27},\rep{27},~\rep{1}) \\
H_4 ~\sim~ (\rep{27},\overline{\rep{27}},~\rep{1}) \nonumber
\eea

The irrep $\rep{27}$ contains a singlet of $SO(10)$, but no singlet
under $SU(6) \times SU(2)$, 
$$
E_6 ~\supset~ SU(6) \times SU(2)~ ; \qquad
\rep{27}~=~(\overline{\rep{6}},\rep{2}) \oplus (\rep{15},\rep{1})
$$
nor under $SU(5) \times SU(2)$ as we have
$$
SU(6) ~\supset~ SU(5) \times U(1)~ ; \qquad
\rep{6}~=~\rep{5}_1 \oplus \rep{1}_{-5}~ ; \qquad
\rep{15}~=~\rep{5}_{-4} \oplus \rep{10}_2
$$
However, we can see that it contains two distinct singlets of $SU(5)$.
The way to understand this is the following. By making suitable
linear combinations, one is a singlet under $SO(10)$, and the other one
is singlet under $SU(5) \times U(1)_V$ but not under $SO(10)$.
We will designate these singlets by $\rep{1}_{(0,4)}$ and
$\rep{1}_{(5,1)}$ respectively. They are distinguished by
their charges $(v,v')$ under $U(1)_V$ and $U(1)_{V'}$.

Similarly, the irrep $\rep{78}$ decomposes itself as
$$
E_6 ~\supset~ SO(10) \times U(1)_{V'}~ ; \qquad
\rep{78} ~=~ \rep{45}_0 \oplus \rep{16}_{-3}
\oplus \overline{\rep{16}}_3 \oplus \rep{1}_0
$$
$$
E_6 ~\supset~ SU(6) \times SU(2)~ ; \qquad~~
\rep{78} ~=~ (\rep{1},\rep{3}) \oplus (\rep{35},\rep{1}) 
\oplus (\rep{20},\rep{2})
$$
Therefore, it contains one singlet of $SO(10)$ but no singlet under
$SU(6) \times SU(2)$. Using the branching rules
$$
SU(6) ~\supset~ SU(5) \times U(1)~ ; \qquad
\rep{20}~=~\rep{10}_{-3} \oplus \overline{\rep{10}}_3
$$
$$
~~~~~~~~~~~~~~~~~~~~~~~~~~~~~~~~~~~~~~~~~~~~~~~~\qquad
\rep{35}~=~\rep{1}_0 \oplus \rep{5}_6 \oplus \overline{\rep{5}}_{-6}
\oplus \rep{24}_0~,
$$
we further see that $\rep{78}$ does contain a singlet under
$SU(5) \times SU(2)$. We can also see that it contains four singlets 
under $SU(5)$ alone, which is in agreement with what is obtained
using the chain of maximal subalgebras~(\ref{chain1}), because
$$
SO(10) ~\supset~ SU(5) \times U(1)_V~; \qquad
\rep{16} ~=~ \rep{10}_1 \oplus \overline{\rep{5}}_{-3} 
\oplus \rep{1}_5
$$
$$
~~~~~~~~~~~~~~~~~~~~~~~~~~~~~~~~~~~~~~~~~~~~~~~~\qquad
\rep{45} ~=~ \rep{24}_0 \oplus \rep{10}_{-4} \oplus 
\overline{\rep{10}}_4+\rep{1}_0
$$
Two of these singlets have a distinct non-zero charge $V$, they 
are designated by $\rep{1}_{(5,3)}$ and $\rep{1}_{(5,-3)}$, following
our notation. The two remaining singlets have zero $V$ and $V'$
charges. By making suitable linear combinations, one is 
singlet under $SO(10)$, and the other one is singlet under
$SU(5) \times SU(2)$. We will designate them collectively
by $\rep{1}_{(0,0)}$, which is sufficient for the discussion of 
$U(1)$ factors. The implications of an extra $SU(2)$ factor in the
framework of Froggatt and Nielsen is beyond the scope of the 
present paper, and will be described elsewhere.

Following our method, a singlet under the diagonal $SU(5)$ 
will be associated with a broken direction in a six-dimensional space,
$\vec{v}=(a,b,c,a',b',c')$ corresponding to the combination
$a V^\alpha + b V^\beta + c V^\gamma + 
a' V'^\alpha + b' V'^\beta + c' V'^\gamma$. 
Many more possibilities of $SU(5)$ singlet directions arise in this case.
Lists of inequivalent singlets (under permutation
of the indices $\alpha$, $\beta$, $\gamma$) and their associated broken
directions are given in the Tables~\ref{sing1} to~\ref{sing3}.

\section{Physical Implications}

\subsection{More on anomalies}

As noticed earlier, singlets that leave the family-blind 
combinations unbroken are of particular interest. 
They trigger a symmetry breaking in which the $U(1)$ symmetries
are factorized into a family-blind part $X$ and a 
family dependent part $Y_F$ which is traceless over the family index.
This decomposition, in turn, enables a "multi-layered" anomaly 
cancellation, which is a central ingredient in the construction
of Froggatt-Nielsen-type models. 

The first layer is given by the cancellation of all the possible
anomalies involving only Standard Model groups over the fermion
content of the Standard Model. This "accidental" cancellation
is now understood in terms of embeddings, because the fermions
of the Standard Model have the right quantum numbers to fit
into the representations $\rep{10}$ and $\overline{\rep{5}}$ of
$SU(5)$, which in turn, complete the $\rep{16}$ of $SO(10)$ when
a right-handed neutrino, singlet under $SU(5)$, is added.

In the second layer, there will be mixed anomalies between $X$ or 
$Y_F$ and Standard Model groups $G_{SM}$. Because the family charge 
$Y_F$ commutes with $SU(5)$ in our framework, it is sufficient
to consider the mixed anomalies between $X$ or $Y_F$ and $SU(5)$.
Anomaly coefficients with a single $SU(5)$, namely 
$(X \, X \, SU(5))$, $(X \, Y_F \, SU(5))$ and $(Y_F \, Y_F \, SU(5))$, 
vanish over any representation of $SU(5)$.
The anomaly coefficient $(Y_F \, SU(5) \, SU(5))$ vanishes over
the family index because $Y_F$ is traceless.
The anomaly coefficients $(X \, SU(5) \, SU(5))$ differ from zero --
$X$ is anomalous -- but are family-independent.
In the low-energy point of view of effective theories of the
Froggatt-Nielsen type, they are cancelled through a dimension five term
at the cut-off, in the so-called Green-Schwarz anomaly cancellation 
mechanism~\cite{GSM}.

In the third layer, the remaining anomaly coefficients do not involve
$G_{SM}$ (or $SU(5)$). The anomalies $(X \, Y_F \, Y_F)$ and
$(X \, X \, X)$ can also be compensated by the Green-Schwarz
mechanism. The coefficient $(Y_F \, X \, X)$ vanish because
$Y_F$ is traceless, so that the only non-vanishing coefficient
that needs to be compensated is $(Y_F \, Y_F \, Y_F)$.

This last cancellation can be achieved by adding new matter fields
which don't carry any Standard Model charge. 
Their presence in the effective theory is solely for the sake of 
anomaly cancellation, and does not modify the observable
phenomenology. 

Of course, in the top-down approach used in our framework, we
know how the question of anomaly cancellation is resolved.
The fermion content of the Standard Model is upgraded
to the $\rep{27}$ of $E_6$ by the addition of vector-like
matter under $SU(5)$, $\rep{5} \oplus \overline{\rep{5}}$,
and two singlets. Therefore, it is not surprising that
the theory looks anomalous when only the Standard Model 
fermions are taken into account! 

\subsection{Predictions for fermion masses and mixings}

Among the possible linear combinations of $V$ and $V'$,
$V^\alpha + V^\beta + V^\gamma$ and $V'^\alpha + V'^\beta + V'^\gamma$
are family-blind. 
Therefore, at least three singlets are needed in order to fix
completely the traceless family symmetry $Y_F$.
If this is the case, the interfamily mass ratios are completely
determined by the family charges $Y_F$ of the fermions, independently
of the dynamics which further breaks the family symmetry.

Let us briefly recall how the family symmetry is related to fermion mass
hierarchies. To fix our notation, we will use the supersymmetric setup of 
Ref.~\cite{FM,FS}. The Yukawa couplings for quarks and charged leptons
stem from invariants in the superpotential of the form
$Q_i \overline{u}_j H_u$, $Q_i \overline{d}_j H_d$, $L_i \overline{e}_j H_d$,
where $i$ and $j$ are family indices.
For neutrinos, after the see-saw mechanism has taken place, the 
effective Yukawa coupling is given by the quartic invariant $L_i L_j H_u^2$.
In the presence of a family symmetry, these invariants can appear
in the superpotential only if they are not charged under the extra
family symmetries. If they are charged, they can still appear, but only
as higher dimensional operators 
$$
\tilde{I} ~=~ I \left(\frac{\theta}{M} \right)^n~,
$$ 
where $I$ is a MSSM invariant, $\theta$ is an order parameter, singlet 
under $G_{SM}$ but charged under $Y_F$, and $M$ is the cut-off scale,
usually taken as the string scale. The operator $\tilde{I}$ has to
be gauge invariant. The families symmetries are spontaneously broken when 
the $\theta$ fields get a vev, yielding the expansion parameter
$$
\lambda ~=~ \frac{<\theta>}{M}
$$
As a result, the Yukawa coupling corresponding to the invariant $I_{ij}$ 
will be suppressed by a power $n_{ij}$ which is related to its family 
charge $Y_F$
\be
n_{ij}=-Y_F(I_{ij})+{\it cst.}
\ee
where the family-independent constant arises because of the anomalous
charge $X$. We can notice that these powers obey the sum rule
$$
n_{ii} + n_{jj} ~=~ n_{ij} + n_{ji} ~.
$$

In our framework, the family symmetry  depends only upon $V$ and $V'$,
and therefore commutes with $SU(5)$. The structure of the mass
matrices is then determined by the charges $-Y_F(\rep{10})$ and 
$-Y_F(\rep{\overline{5}})$, reordered and normalized to the
heaviest family
\bean
Q ~,~ \bar{u} ~,~ \bar{e} ~&\longrightarrow&~ -Y_F(\rep{10}) \\
L ~,~ \bar{d} ~&\longrightarrow&~ -Y_F(\rep{\overline{5}})
\eean

We have taken all the possible sets of three singlets that determine 
the family symmetry $Y_F$, and derived the corresponding Yukawa structure.
However, thousands of different patterns can emerge. 
To reduce this number, we chose to enforce a phenomenological constraint.
To make it strong and reliable, only the heavier fermions from the 
second and third families are involved. 

The measured fermion mass ratios scale as
\be
\frac{m_c}{m_t} ~\sim~ \lambda^4~; \qquad
\frac{m_s}{m_b} ~\sim~ \lambda^2~; \qquad
\frac{m_\mu}{m_\tau} ~\sim~ \lambda^2  
\ee 
where the expansion parameter $\lambda$ is of the order of the
Cabibbo angle $\lambda _c \simeq 0.22$. 
Therefore, we can use the constraint derived from the relations
\be
\frac{m_c}{m_t} ~\sim~ \left( \frac{m_s}{m_b} \right)^2
~\sim~ \left( \frac{m_\mu}{m_\tau} \right)^2
\label{eq:constraint}
\ee
After noticing that Eq.~(\ref{eq:constraint}) is compatible with a family
symmetry commuting with $SU(5)$, we see that it translates into
\be
Y_F(\overline{\rep{5}}_2) ~=~ Y_F(\overline{\rep{5}}_3)
\label{YFeq}
\ee
This relation, in turn, implies a large mixing angle for the
atmospheric neutrinos, as it is indeed observed~\cite{SKatm} !

It turns out that the constraint~(\ref{YFeq}) restricts severely the
number of possible patterns. 
Four 'scenarios' were considered, in which
the order parameter that is spontaneously broken transforms
as $H_1$, $H_2$, $H_4$ and as $H_1$ or $H_3$ in the last 
one (see Eq.~(\ref{eq:hach})).
We find that a bi-chiral or a bi-vectorial order parameter
does not lead to a family symmetry with physical interest.
The family charges for a tri-chiral or tri-vectorial order
parameter that survive from the constraint~(\ref{YFeq}) are listed
in Tables~6~and~7. 
The interesting charges that give rise to a phenomenology 
compatible with all present data on quarks, charged leptons and 
neutrinos masses and mixings have been underlined. 
It can be noticed that all of them can be obtained
with an order parameter transforming like $H_2$. 

Most of these charges are only in weak agreement with the data.
The mismatch lies in the predicted masses for the first 
generation. A light up quark mass is in conflict with
the derivation of the correct CKM matrix.
A further conflict comes from the neutrinos sector,
where a mild $\Delta m^2$ hierarchy, and a large solar
mixing angle induce a heavier electron. 

In the following, we analyze in more details three of these 
possible charge assignments.

\subsubsection*{\underline{Model A}}

The charges $-Y_F(\rep{10}) \sim (2,1,0)$ and 
$-Y_F(\overline{\rep{5}}) \sim (0,0,0)$ with expansion parameter
$\lambda \sim \lambda _c^2$ give rise to the so-called
anarchical model~\cite{berger}, where hierarchical and mixing
structure is totally absent from the neutrino mass matrix.
The observed hierarchy between the $\Delta m^2_\oplus$ for atmospheric 
neutrinos and the $\Delta m^2_\odot$ for solar neutrinos 
(for the LMA solution~\cite{solnu})
\be
\frac{\Delta m^2_\odot}{\Delta m^2_\oplus} ~\simeq~ 10^{-2}
\label{eq:nuhier}
\ee 
can still be obtained but resides in Nature's choice of the
prefactor coefficients. Large mixing angles for solar and 
atmospheric neutrinos are natural in this context, but
the small value of the CHOOZ~\cite{CHOOZ} mixing angle is problematic.
The model is otherwise in fair agreement with phenomenology.
The predicted mass ratios are
\bean
\frac{m_u}{m_t} ~\sim~ \lambda _c^8~; \qquad
\frac{m_c}{m_t} ~\sim~ \lambda _c^4 \\
\frac{m_d}{m_b} ~\sim~ \lambda _c^4~; \qquad
\frac{m_s}{m_b} ~\sim~ \lambda _c^2 \\
\frac{m_e}{m_\tau} ~\sim~ \lambda _c^4~; \qquad
\frac{m_\mu}{m_\tau} ~\sim~ \lambda _c^2
\eean
However, we notice that the electron mass comes out too heavy 
compared to the $\tau$ mass. A smaller ratio 
$$
\frac{m_e}{m_\tau} ~\sim~ \lambda _c^{5-6}
$$ 
would be in better agreement with the measured masses.
Moreover, some stretching in the prefactor coefficients is 
needed to reconcile the model with the structure of the 
CKM matrix~\cite{berger}.  

\subsubsection*{\underline{Model B}}

The charges $-Y_F(\rep{10}) \sim (3,2,0)$ and 
$-Y_F(\overline{\rep{5}}) \sim (2,0,0)$ with $\lambda \simeq \lambda _c$ 
enable to reproduce the CKM matrix. The predictions for quarks and 
charged leptons are
\bean
\frac{m_u}{m_t} ~\sim~ \lambda _c^6~; \qquad
\frac{m_c}{m_t} ~\sim~ \lambda _c^4 \\
\frac{m_d}{m_b} ~\sim~ \lambda _c^4~; \qquad
\frac{m_s}{m_b} ~\sim~ \lambda _c^2 \\
\frac{m_e}{m_\tau} ~\sim~ \lambda _c^4~; \qquad
\frac{m_\mu}{m_\tau} ~\sim~ \lambda _c^2
\eean
with a heavier up quark. In the neutrino sector, the resulting
$\Delta m^2$ hierarchy is in agreement with Eq.~(\ref{eq:nuhier}),
the CHOOZ mixing angle is suppressed by a factor $\lambda _c^2$,
but the solar mixing angle also turns out to be naturally small,
which is less satisfying in view of the recent solar neutrino 
data~\cite{solnu}, and the KamLAND result~\cite{kamland}. 

A phenomenologically better charge assignment would be
$-Y_F(\rep{10}) \sim (3,2,0)$ and $-Y_F(\overline{\rep{5}}) \sim (1,0,0)$
It can been incorporated in a consistent way in a FN model~\cite{FS}, 
but could not be reproduced by symmetry breaking in the present approach.

\subsubsection*{\underline{Model C}}

Finally, the charges $-Y_F(\rep{10}) \sim (5,2,0)$ and 
$-Y_F(\overline{\rep{5}}) \sim (1,0,0)$ also with 
$\lambda \simeq \lambda _c$ can be in fairly good
agreement with the observations. 
The predicted mass ratios are
\bean
\frac{m_u}{m_t} ~\sim~ \lambda _c^{10}~; \qquad
\frac{m_c}{m_t} ~\sim~ \lambda _c^4 \\
\frac{m_d}{m_b} ~\sim~ \lambda _c^6~; \qquad
\frac{m_s}{m_b} ~\sim~ \lambda _c^2 \\
\frac{m_e}{m_\tau} ~\sim~ \lambda _c^6~; \qquad
\frac{m_\mu}{m_\tau} ~\sim~ \lambda _c^2
\eean
We notice that the electron to $\tau$ mass ratio
have the correct order of magnitude, but the up and the down quarks
appear a little light. 
However, it has been pointed out that non perturbative QCD effects 
can also contribute to the masses of the lightest quarks at low-energy,
so that the actual masses and the mixings in the CKM matrix can
be recovered~(see~\cite{BNS} for example).

\bt
\caption{\label{sing1} Singlets of $SU(5)$ in the product
$\rep{27}^\alpha \otimes \rep{27}^\beta \otimes \rep{27}^\gamma$ 
and associated broken directions $(a,b,c,a',b',c')$.
Singlets obtained by permutation of the indices $\alpha$,
$\beta$, $\gamma$ are omitted.}
\bc
\begin{tabular}{|c|c|c|}
\hline
&&\\
$SU(5)$ Singlet in & Unbroken & Broken\\
$\rep{27} \otimes \rep{27} \otimes \rep{27}$ &
non-Abelian Group & Direction\\&&\\
\hline \hline &&\\
$\rep{1}^\alpha _{(0,4)} \otimes \rep{1}^\beta _{(0,4)}
\otimes \rep{1}^\gamma _{(0,4)}$ & 
$SO(10)^\alpha \times SO(10)^\beta \times SO(10)^\gamma$ & 
$(0,0,0,1,1,1)$\\&&\\
$\rep{1}^\alpha _{(5,1)} \otimes \rep{1}^\beta _{(0,4)}
\otimes \rep{1}^\gamma _{(0,4)}$ & 
$SU(5)^\alpha \times SO(10)^\beta \times SO(10)^\gamma$ & 
$(5,0,0,1,4,4)$\\&&\\
$\rep{1}^\alpha _{(5,1)} \otimes \rep{1}^\beta _{(5,1)}
\otimes \rep{1}^\gamma _{(0,4)}$ &
$SU(5)^\alpha \times SU(5)^\beta \times SO(10)^\gamma$ & 
$(5,5,0,1,1,4)$\\&&\\
$\rep{1}^\alpha _{(5,1)} \otimes \rep{1}^\beta _{(5,1)}
\otimes \rep{1}^\gamma _{(5,1)}$ & 
$SU(5)^\alpha \times SU(5)^\beta \times SU(5)^\gamma$ & 
$(5,5,5,1,1,1)$\\&&\\
\hline &&\\

$\rep{1}^{\alpha+\beta}_{(0,-4,4,0)} 
\otimes \rep{1}^\gamma _{(0,4)}$ & 
$SU(5)^{\alpha+\beta} \times SO(10)^\gamma$ & 
$(1,-1,0,-1,-1,2)$\\&&\\
$\rep{1}^{\alpha+\beta}_{(-5,-1,-1,3)} 
\otimes \rep{1}^\gamma _{(0,4)}$ & 
$SU(5)^{\alpha+\beta} \times SO(10)^\gamma$ & 
$(-3,-2,0,1,-2,4)$\\&&\\
$\rep{1}^{\alpha+\beta}_{(-5,-1,-1,3)} 
\otimes \rep{1}^\gamma _{(5,1)}$ & 
$SU(5)^{\alpha+\beta} \times SU(5)^\gamma$ & 
$(-3,-2,5,1,-2,1)$\\&&\\
$\rep{1}^{\alpha+\beta}_{(0,-4,4,0)} 
\otimes \rep{1}^\gamma _{(5,1)}$ & 
$SU(5)^{\alpha+\beta} \times SU(5)^\gamma$ & 
$(2,-2,5,-2,-2,1)$\\&&\\
\hline &&\\

$\rep{1}^{\alpha+\beta+\gamma}_{(-5,3,4,0,4,0)}$ 
&& ${(1,-3,-3,1,1,1)}$\\&&\\
$\rep{1}^{\alpha+\beta+\gamma}_{(0,0,-5,3,-4,0)}$ 
&& ${(-3,2,1,1,-2,1)}$\\& 
$SU(5)^{\alpha+\beta+\gamma}$ & \\
$\rep{1}^{\alpha+\beta+\gamma}_{(0,0,0,0,3,3)}$ 
&& ${(1,1,-2,1,1,-2)}$\\&&\\
$\rep{1}^{\alpha+\beta+\gamma}_{(5,-3,0,0,1,-3)}$ 
&& ${(2,2,1,-2,-2,1)}$\\&&\\
\hline
\end{tabular}
\ec
\et

\bt
\caption{\label{sing2} Singlets of $SU(5)$ in the product
$\rep{78}^\alpha \otimes \rep{78}^\beta \otimes \rep{78}^\gamma$ 
and associated broken directions. 
Singlets obtained by permutation of the indices $\alpha$,
$\beta$, $\gamma$  and conjugate singlets (with the sign of all charges
reversed) are omitted, and $G$ stands for either $SO(10)$ or
$SU(5) \times SU(2)$.}
\bc
\begin{tabular}{|c|c|c|}
\hline &&\\
$SU(5)$ Singlet in & Unbroken & Broken\\
$\rep{78} \otimes \rep{78} \otimes \rep{78}$ & 
non-Abelian Group & Direction\\&&\\
\hline \hline &&\\

$\rep{1}^\alpha _{(5,-3)} \otimes \rep{1}^\beta _{(5,-3)}
\otimes \rep{1}^\gamma _{(5,-3)}$ & 
$SU(5)^\alpha \times SU(5)^\beta \times SU(5)^\gamma$ & 
$(5,5,5,-3,-3,-3)$\\&&\\
$\rep{1}^\alpha _{(5,-3)} \otimes \rep{1}^\beta _{(5,-3)}
\otimes \rep{1}^\gamma _{(-5,3)}$ & 
$SU(5)^\alpha \times SU(5)^\beta \times SU(5)^\gamma$ & 
$(5,5,-5,-3,-3,3)$\\&&\\
$\rep{1}^\alpha _{(5,-3)} \otimes \rep{1}^\beta _{(5,-3)}
\otimes \rep{1}^\gamma _{(0,0)}$ & 
$SU(5)^\alpha \times SU(5)^\beta \times G^\gamma$ & 
$(5,5,0,-3,-3,0)$\\&&\\
$\rep{1}^\alpha _{(5,-3)} \otimes \rep{1}^\beta _{(-5,3)}
\otimes \rep{1}^\gamma _{(0,0)}$ & 
$SU(5)^\alpha \times SU(5)^\beta \times G^\gamma$ & 
$(5,-5,0,-3,3,0)$\\&&\\
$\rep{1}^\alpha _{(5,-3)} \otimes \rep{1}^\beta _{(0,0)}
\otimes \rep{1}^\gamma _{(0,0)}$ & 
$SU(5)^\alpha \times G^\beta \times G^\gamma$ & 
$(5,0,0,-3,0,0)$\\&&\\
\hline &&\\

$\rep{1}^{\alpha+\beta}_{(0,0,6,6)} 
\otimes \rep{1}^\gamma _{(0,0)}$ && $(1,-1,0,1,-1,0)$\\&&\\
$\rep{1}^{\alpha+\beta}_{(0,0,-8,0)} 
\otimes \rep{1}^\gamma _{(0,0)}$ && $(1,-1,0,0,0,0)$\\&
$SU(5)^{\alpha+\beta} \times G^\gamma$&\\
$\rep{1}^{\alpha+\beta}_{(0,0,2,-6)} 
\otimes \rep{1}^\gamma _{(0,0)}$ && $(1,-1,0,-3,3,0)$\\&&\\
$\rep{1}^{\alpha+\beta}_{(-5,3,-3,-3)} 
\otimes \rep{1}^\gamma _{(0,0)}$ && $(-4,-1,0,0,3,0)$\\&&\\
\hline &&\\

$\rep{1}^\alpha _{(0,0)} \otimes \rep{1}^\beta _{(0,0)}
\otimes \rep{1}^\gamma _{(0,0)}$ & 
$G^\alpha \times G^\beta \times G^\gamma$ & 
None\\&&\\
$\rep{1}^{\alpha+\beta}_{(0,0,0,0)} 
\otimes \rep{1}^\gamma _{(0,0)}$ & 
$SU(5)^{\alpha+\beta} \times G^\gamma$ & 
None\\&&\\
$\rep{1}^{\alpha+\beta+\gamma} _{(0,0,0,0,0,0)}$ & 
$SU(5)^{\alpha+\beta+\gamma}$ & 
None\\&&\\

\hline
\end{tabular}
\ec
\et

\bt
\caption{\label{sing3} Singlets of $SU(5)$ in the product
$\rep{78}^\alpha \otimes \rep{78}^\beta \otimes \rep{78}^\gamma$ 
and associated broken directions (continued).}
\bc
\begin{tabular}{|c|c|c|}
\hline
&&\\
$SU(5)$ Singlet in & Unbroken & Broken\\
$\rep{78} \otimes \rep{78} \otimes \rep{78}$ & 
non-Abelian Group & Direction\\&&\\
\hline \hline &&\\

$\rep{1}^{\alpha+\beta}_{(0,0,6,6)} 
\otimes \rep{1}^\gamma _{(5,-3)}$ && $(3,-3,5,3,-3,-3)$\\&&\\
$\rep{1}^{\alpha+\beta}_{(0,0,-8,0)} 
\otimes \rep{1}^\gamma _{(5,-3)}$ && $(-4,4,5,0,0,-3)$\\&&\\
$\rep{1}^{\alpha+\beta}_{(0,0,2,-6)} 
\otimes \rep{1}^\gamma _{(5,-3)}$ &
~~~$SU(5)^{\alpha+\beta} \times SU(5)^\gamma$~~~
& $(1,-1,5,-3,3,-3)$\\&&\\
~~$\rep{1}^{\alpha+\beta}_{(-5,3,-3,-3)} 
\otimes \rep{1}^\gamma _{(5,-3)}$~~ && $(-4,-1,5,0,3,-3)$\\&&\\
$\rep{1}^{\alpha+\beta}_{(-5,3,-3,-3)} 
\otimes \rep{1}^\gamma _{(-5,3)}$ && $(-4,-1,-5,0,3,3)$\\&&\\
\hline &&\\

$\rep{1}^{\alpha+\beta+\gamma}_{(10,6,0,0,-1,3)}$ 
&& ${(3,3,4,3,3,0)}$\\&&\\
$\rep{1}^{\alpha+\beta+\gamma}_{(5,9,0,0,4,0)}$ 
&& ${(3,3,-1,3,3,3)}$\\&&\\
$\rep{1}^{\alpha+\beta+\gamma}_{(5,-3,0,0,7,3)}$ 
& $SU(5)^{\alpha+\beta+\gamma}$
& ${(4,4,-3,0,0,-3)}$\\&&\\
$\rep{1}^{\alpha+\beta+\gamma}_{(-5,3,0,0,2,6)}$ 
&& ${(-1,-1,-3,3,3,-3)}$\\&&\\
$\rep{1}^{\alpha+\beta+\gamma}_{(0,0,5,-3,7,3)}$ 
&& ${(4,-1,-3,0,3,-3)}$\\&&\\

\hline
\end{tabular}
\ec
\et



\begin{table}[t]
\bc
\begin{tabular}{cccc}
\mco{4}{l}{Table~6 : Family charges $(-Y_F(\rep{10}),-Y_F(\overline{\rep{5}}))$
normalized to the heaviest}\\
\mco{4}{l}{~~~~~~~~~~~~~family (third family), obtained from $H_1 \sim 
(\rep{27},\rep{27},\rep{27})$.}\\
\hline\\&&&\\
(1, 1, 0), (0, 0, 0) & (1, 1, 0), (1, 0, 0) & 
(1, 1, 0), (2, 0, 0) & (1, 1, 0), (5, 0, 0)\\&&&\\
(1, 1, 0), (8, 0, 0) & (2, 2, 0), (1, 0, 0) &
(2, 2, 0), (7, 0, 0) & \underline{(3, 1, 0), (1, 0, 0)}\\&&&\\
(3, 1, 0), (4, 0, 0) & \underline{(3, 2, 0), (2, 0, 0)} & 
\underline{(4, 3, 0), (1, 0, 0)} & (5, 2, 0), (4, 0, 0)
\end{tabular}
\ec
\et

\begin{table}[t]
\bc
\begin{tabular}{cccc}
\mco{4}{l}{Table~7 : Family charges $(-Y_F(\rep{10}),-Y_F(\overline{\rep{5}}))$
normalized to the heaviest}\\
\mco{4}{l}{~~~~~~~~~~~~~family (third family), 
obtained from $H_2 \sim (\rep{78},\rep{78},\rep{78})$.}\\
\hline\\&&&\\
(1, 1, 0), (2, 0, 0) & (1, 1, 0), (5, 0, 0) &
\underline{(2, 1, 0), (0, 0, 0)} & (2, 1, 0), (3, 0, 0)\\ &&&\\
(2, 2, 0), (1, 0, 0) & \underline{(3, 1, 0), (1, 0, 0)} & 
(3, 1, 0), (4, 0, 0) & (3, 1, 0), (16, 0, 0)\\&&&\\
\underline{(3, 2, 0), (2, 0, 0)} & (3, 2, 0), (5, 0, 0) & 
(4, 1, 0), (2, 0, 0) & (4, 1, 0), (5, 0, 0)\\&&&\\
\underline{(4, 3, 0), (1, 0, 0)} & (4, 3, 0), (10, 0, 0) & 
(5, 1, 0), (3, 0, 0) & \underline{(5, 2, 0), (1, 0, 0)}\\ &&&\\
(5, 2, 0), (4, 0, 0) & (5, 5, 0), (7, 0, 0) &
(6, 2, 0), (5, 0, 0) & (6, 5, 0), (8, 0, 0)\\ &&&\\
(7, 4, 0), (5, 0, 0) & (8, 5, 0), (1, 0, 0) & 
(9, 5, 0), (2, 0, 0) &
\end{tabular}
\ec
\et

\section{Conclusions}

The possibility that the fundamental gauge group at very high
scales appears replicated in several copies, as inspired
by the brane world, opens up new possibilities to understand
the patterns in the low energy world.
We have studied how the family mass hierarchy problem can be
elucidated in this context through a group-theoretical approach.
Imprints of symmetry breakings were recognized and systematically
analyzed, although the dynamics which triggers the symmetry 
breaking is beyond the scope of the present work.

The realistic scheme considered, based on 
$E_6 \times E_6 \times E_6 \rightarrow SU(5)$
shows interesting features. 
Although many different family symmetries can survive
after symmetry breaking, a simple and reliable phenomenological
constraint killed most of them. Moreover, only a few possibilities
lead to mass patterns in accordance with observations.
So it might be that our world has something {\it exceptional}
rather than {\it generic}, that very particular dynamical
conditions triggered such a symmetry breaking.
Or it might also be that the true mechanism that Nature choose
to order fermion masses is totally different.

However, it is worth pointing that this approach \textbf{\emph{does}}
give rise to patterns in agreement with our world.
Moreover, it appears that this path, for reasons explained
throughout this paper, necessarily leads to the consideration
of exceptional algebras. So Nature might indeed be {\it exceptional}...
Anyhow, the idea that mass could be partially treated as a quantum
number is a very attractive scheme that can help us organize and 
understand the legacy of the Standard Model.

\section*{Acknowledgments}

This work is supported by the United States Department of Energy
under grant DE-FG02-97ER41029.


\end{document}